\documentclass[apl,amsmath,reprint,floatfix,a4paper]{revtex4-1}

\usepackage[T1]{fontenc}
\usepackage{etex}
\usepackage{geometry}
\usepackage{color,tabularx,colortbl}

\usepackage{setspace}
\usepackage{pgfplots}
\usepackage{booktabs}
\usepackage{multirow}
\usepackage{tikz, tikz-3dplot}
\tikzset{>=latex}
\usepackage[percent]{overpic}
\pgfplotsset{compat=1.13}
\usepgfplotslibrary{external} 
\usepgfplotslibrary{fillbetween}
\usepgfplotslibrary{colorbrewer}
\usetikzlibrary{pgfplots.groupplots,arrows,decorations.markings}
\usetikzlibrary{shapes.geometric,calc,patterns}
\usepackage[strict]{changepage}

\usepackage{sidecap}
\sidecaptionvpos{figure}{t}

\usepackage{todonotes}
\makeatletter
\renewcommand{\todo}[2][]{\tikzexternaldisable\@todo[#1]{#2}\tikzexternalenable}
\renewcommand{\missingfigure}[2][]{\tikzexternaldisable\@missingfigure[#1]{#2}\tikzexternalenable}
\makeatother

\usepackage{dcolumn}
\newcolumntype{d}[1]{D{.}{.}{#1}}

\pgfmathsetmacro{\muB}{9.27400968e-24}
\pgfmathsetmacro{\HaniA}{2.0*8.0725/(2.217*9.27400968)}
\pgfmathsetmacro{\HaniB}{2.0*8.0725/(7.63*9.27400968)}
\pgfmathsetmacro{\sigmaGaussFWHMlarge}{1.69864360058}
\pgfmathsetmacro{\sigmaGaussFWHMsmall}{0.849321800288}
\pgfmathsetmacro{\Tmin}{270}
\pgfmathsetmacro{\TClarge}{644.326893845}
\pgfmathsetmacro{\TCsmall}{536.939078204}

\pgfmathsetmacro{\Tfreeze}{364}
\pgfmathsetmacro{\Tjump}{442.335666891}
\pgfmathsetmacro{\tfreeze}{0.483233608052}
\pgfmathsetmacro{\tjump}{0.535762818744}
\pgfmathsetmacro{\Tc}{537}

\definecolor{android_blue}{RGB}{51,181,229}
\definecolor{android_dark_blue}{RGB}{0,153,204}
\definecolor{android_pink}{RGB}{170,102,204}
\definecolor{android_purple}{RGB}{156,39,176}
\definecolor{android_dark_pink}{RGB}{153,51,204}
\definecolor{android_green}{RGB}{153,204,0}
\definecolor{android_dark_green}{RGB}{102,153,0}
\definecolor{android_orange}{RGB}{255,152,0}
\definecolor{android_dark_orange}{RGB}{255,152,0}
\definecolor{android_red}{RGB}{255,68,68}
\definecolor{android_dark_red}{RGB}{204,0,0}
\definecolor{android_pink}{RGB}{156,39,176}
\definecolor{android_grey}{RGB}{158,158,158}

\pgfplotsset{grid style={dashed,android_grey,opacity=0.5}}
	  
\pgfplotscreateplotcyclelist{peak_temp}{
	  {color=android_green,line width=0.5pt,mark size=2pt,mark options={line width=0.75pt},line join=round},
	  {color=android_red,line width=0.5pt,mark size=2pt,mark options={line width=0.75pt},line join=round},
	  {color=android_blue,line width=0.5pt,mark size=2pt,mark options={line width=0.75pt},line join=round},
	  {color=android_pink,line width=0.5pt,mark size=2pt,mark options={line width=0.75pt},line join=round},
	  {color=android_orange,line width=0.5pt,mark size=2pt,mark options={line width=0.75pt},line join=round},
	  {color=black,line width=0.5pt,mark size=2pt,mark options={line width=0.75pt},line join=round}}

\begin{document}

\title{Stochastic ferrimagnetic Landau-Lifshitz-Bloch equation\\for finite magnetic structures} 

\author{Christoph Vogler}
\email{christoph.vogler@univie.ac.at}
\affiliation{Faculty of Physics, University of Vienna, Boltzmanngasse 5, 1090 Vienna, Austria}

\author{Claas Abert}
\author{Florian Bruckner}
\author{Dieter Suess}
\affiliation{Christian Doppler Laboratory for Advanced Magnetic Sensing and Materials, Faculty of Physics, University of Vienna, Boltzmanngasse 5, 1090 Vienna, Austria}

\begin{abstract}
Precise modeling of the magnetization dynamics of nanoparticles with finite size effects at fast varying temperatures is a computationally challenging task. Based on the Landau-Lifshitz-Bloch (LLB) equation we derive a coarse grained model for disordered ferrimagnets, which is both fast and accurate. First, we incorporate stochastic fluctuations to the existing ferrimagnetic LLB equation. Further, we derive a thermodynamic expression for the temperature dependent susceptibilities, which is essential to model finite size effects. Together with the zero field equilibrium magnetization the susceptibilities are used in the stochastic ferrimagnetic LLB to simulate a $5\times10$\,nm$^2$ ferrimagnetic GdFeCo particle with 70\,\% FeCo and 30\,\% Gd under various external applied fields and heat pulses. The obtained trajectories agree well with those of an atomistic model, which solves the stochastic Landau-Lifshitz-Gilbert equation for each atom. Additionally, we derive an expression for the intergrain exchange field which couple the ferromagnetic sublattices of a ferrimagnet. A comparison of the magnetization dynamics obtained from this simpler model with those of the ferrimagnetic LLB equation shows a perfect agreement.
\end{abstract}

\maketitle 

\pgfplotsset{colormap/RdBu-9}

\section{Introduction}
\label{sec:intro}
The calculation of the magnetization dynamics of large systems under the influence of fast varying temperatures is of great interest from both the scientific and the technological perspective. Heat-assisted magnetic recording (HAMR)~\cite{mayer_curiepoint_1958,mee_proposed_1967,guisinger_thermomagnetic_1971,kobayashi_thermomagnetic_1984,rottmayer_heat-assisted_2006} should be mentioned first and foremost here. Despite the computing power of modern supercomputers, coarse-grained models are needed to manage the computational effort created by such complex systems. The development of the Landau-Lifshitz-Bloch (LLB) equation for pure ferromagnets by Garanin~\cite{garanin_fokker-planck_1997} and the subsequent improvements~\cite{kazantseva_towards_2008,evans_stochastic_2012,volger_llb} paved the way to make concrete design proposals for real HAMR devices~\cite{zhu2013understanding,vogler_heat-assisted_2016,zhu_correcting_2017,vogler_efficiently_2017,vogler_noise_2017}. 

Similar to the derivation of the Landau-Lifshitz-Bloch (LLB) equation for pure ferromagnets by Garanin~\cite{garanin_fokker-planck_1997} (see Appendix~\ref{sec:appendix}), Atxitia~et~al.~\cite{atxitia_landau-lifshitz-bloch_2012} have recently shown how the LLB equation can be adapted for disordered ferrimagnets with two sublattices. Before going into detail and presenting extensions to ferrimagnetic LLB equation, we would like to briefly review the results of Ref.~\cite{atxitia_landau-lifshitz-bloch_2012}. The temporal evolution of the reduced magnetization $\boldsymbol{m}_\mathrm{A}=\boldsymbol{M}_\mathrm{A}/M_{\mathrm{A,}0}$ (with $M_{\mathrm{A,}0}$ being the zero temperature sublattice saturation magnetization) of sublattice $\mathrm{A}$ can be calculated per
\begin{eqnarray}
 \label{eq:LLB_ferri}
 \frac{\partial \boldsymbol{m}_\mathrm{A}}{\partial t}=&-& \mu_0\gamma'_\mathrm{A} \left ( \boldsymbol{m}_\mathrm{A} \times \boldsymbol{H}_{\mathrm{eff,}\mathrm{A}} \right )\nonumber \\
 &+& \frac{\mu_0\gamma'_\mathrm{A} \alpha_\mathrm{A}^\parallel}{m_\mathrm{A}^2} \left ( \boldsymbol{m}_\mathrm{A} \cdot \boldsymbol{H}_{\mathrm{eff,}\mathrm{A}} \right )\boldsymbol{m}_\mathrm{A}\nonumber \\
 &-& \frac{\mu_0\gamma'_\mathrm{A} \alpha_\mathrm{A}^\perp}{m_\mathrm{A}^2}\left [ \boldsymbol{m}_\mathrm{A} \times \left ( \boldsymbol{m}_\mathrm{A} \times \boldsymbol{H}_{\mathrm{eff,}\mathrm{A}} \right )\right ],
\end{eqnarray}
where $\alpha_\mathrm{A}^\perp$ and $\alpha_\mathrm{A}^\parallel$ are the perpendicular and the parallel dimensionless damping constants, respectively. $\gamma'_\mathrm{A}$ is the reduced electron gyromagnetic ratio $\gamma'_\mathrm{A}=\gamma_{\mathrm{e}}/(1+\lambda_\mathrm{A}^2)$, which is defined via the coupling parameter $\lambda_\mathrm{A}$ of sublattice A to the heat bath. It is not surprising that Eq.~\ref{eq:LLB_ferri} is of the same form as the ferroLLB equation, because within each sublattice the magnetizations and the field terms are treated with the mean field approximation usually used for ferromagnets.

The effective field $\boldsymbol{H}_{\mathrm{eff},\mathrm{A}}$ of each sublattice is defined per~\cite{atxitia_landau-lifshitz-bloch_2012}
\begin{eqnarray}
 \label{eq:LLB_Hpara}
  \mu_0\boldsymbol{H}_{\mathrm{eff},\mathrm{A}}=&& \mu_0\boldsymbol{H}_{\mathrm{ext}}+\frac{2d_{\mathrm{A}}}{\mu_{\mathrm{A}}}m_{z\mathrm{,A}}\boldsymbol{e}_z\nonumber \\
 &&-\frac{J_{0,\mathrm{A}\mathrm{B}}}{\mu_\mathrm{A} m_\mathrm{A}^2}\left[\boldsymbol{m}_\mathrm{A}\times\left( \boldsymbol{m}_\mathrm{A} \times \boldsymbol{m}_\mathrm{B}\right)\right] \nonumber \\
 &&+ \Lambda_1 \left [1- \left (\frac{\boldsymbol{m}_\mathrm{A} \cdot \boldsymbol{m}_\mathrm{B}}{\boldsymbol{m}_{\mathrm{e},\mathrm{A}} \cdot \boldsymbol{m}_{\mathrm{e},\mathrm{B}}}\right ) ^2 \frac{m_{\mathrm{e},\mathrm{A}}^2}{m_\mathrm{A}^2} \right ] \boldsymbol{m}_\mathrm{A}\nonumber \\
 &&- \Lambda_2 \left ( 1- \frac{m_\mathrm{A}^2}{m_{\mathrm{e},\mathrm{A}}^2} \right )\boldsymbol{m}_\mathrm{A},
\end{eqnarray}
with
\begin{equation}
 \label{eq:lambda1}
 \Lambda_1=\frac{|\boldsymbol{m}_{\mathrm{e},\mathrm{A}} \cdot \boldsymbol{m}_{\mathrm{e},\mathrm{B}}|}{2 m_{\mathrm{e},\mathrm{A}}^2} \frac{|J_{0,\mathrm{A}\mathrm{B}}|}{\mu_\mathrm{A}}.
\end{equation}
and
\begin{equation}
 \label{eq:lambda}
 \Lambda_2=\frac{1}{2 \tilde{\chi}_\mathrm{A}^\parallel} \left( 1+\frac{|J_{0,\mathrm{AB}}|}{\mu_\mathrm{A}} \tilde{\chi}_\mathrm{B}^\parallel \right).
\end{equation}
Here, $\mu_\mathrm{A}$ is the magnetic moment of each spin in sublattice A, $d_{\mathrm{A}}$ is the uniaxial anisotropy energy per spin, $m_{\mathrm{e},\mathrm{A}}$ is the equilibrium magnetization and $\tilde{\chi}_\mathrm{A}^\parallel$ is the longitudinal susceptibility of the sublattice. In sublattice $\mathrm{B}$ the same quantities are defined. In the case of two sublattices with atoms A and B there exist three exchange energies, $J_{\mathrm{A-A}}$, $J_{\mathrm{B-B}}$ and $J_{\mathrm{A-B}}$. The exchange energies in the LLB model depend on the number of nearest neighbors $z$ and on the concentrations $x_\mathrm{A}$ of the atoms. Hence, the exchange energies become $J_{0,\mathrm{A}\mathrm{A}}=zx_\mathrm{A} J_{\mathrm{\mathrm{A}-\mathrm{A}}}$ and $J_{0,\mathrm{B}\mathrm{A}}=zx_\mathrm{A} J_{\mathrm{\mathrm{A}-\mathrm{B}}}$. 

The described formalism was successfully applied in the past~\cite{atxitia_ultrafast_2013,atxitia_controlling_2014,suarez_ultrafast_2015,hinzke_multiscale_2015,atxitia_fundamentals_2017}. Most of these works investigate fast relaxation processes in ferrimagnets and use a simplified or a linearized version of the ferrimagnetic LLB. Due to the deterministic nature of Eq.~\ref{eq:LLB_ferri} all results can be interpreted as ensemble averages. We are interested in the full dynamical response of ferrimagnets with finite size under arbitrary external conditions. In the presence of temperature, this response has a stochastic nature. 

Note, all equations are identical for sublattice B if subscript A is replaced by subscript B. For the sake of clarity we will call the LLB equation for pure ferromagnets ferroLLB (see Appendix~\ref{sec:appendix}) and the LLB equation for ferrimagnets ferriLLB equation in the following.

\section{Extensions to the ferrimagnetic LLB equation}
\subsection{stochastic form}
To account for stochastic fluctuations due to temperature we follow the derivations of Evans~et~al.~\cite{evans_stochastic_2012} for the ferroLLB equation, which lead to a Boltzmann distribution of the magnetization in equilibrium. The basic assumption is that thermal fluctuations can be introduced to the LLB via thermal fields. These fields are uncorrelated in time and space, which means that its components consist of white noise random numbers with zero mean and a variance
\begin{equation}
\label{eq:fluct}
 \left< \xi_{\kappa,i}^\eta(t,\boldsymbol{r})\xi_{\kappa,j}^\eta(t',\boldsymbol{r}') \right>=2D_\kappa^\eta \delta_{ij}\delta(\boldsymbol{r}-\boldsymbol{r}')\delta(t-t'),
\end{equation}
where $i,j$ are the Cartesian components of the thermal field, $\kappa$ is a placeholder for the sublattice type (A or B) and $\eta$ is a placeholder for parallel and perpendicular field components. The four diffusion constants $D_\kappa^\eta$ are to be determined for the specific problem. To achieve this there exist two strategies, one by means of the fluctuation dissipation theorem and one via the Fokker-Planck equation. We will use the latter approach in the following.

In its most general form the LLB equation can be written as a multivariate Langevin equation:
\begin{equation}
 \label{eq:Langevin}
 \frac{dm_{i}}{dt}=a_i(\boldsymbol{m},t)+\sum_{k\eta}b_{ik}^\eta(\boldsymbol{m},t)\xi_k^\eta(t).
\end{equation}
If the vector $a_i(\boldsymbol{m},t)$ and the tensor $b_{ik}(\boldsymbol{m},t)$ are known the corresponding Fokker-Planck (FP) equation can be directly constructed per
\begin{eqnarray}
\label{eq:FP}
 \frac{\partial \rho}{\partial t}=-\sum_{i}\frac{\partial}{\partial m_i}\Bigg[ &\Bigg ( &a_i-\sum_\eta D^\eta \sum_{k} b_{ik}^\eta\sum_{j} \frac{\partial b_{jk}^\eta}{\partial m_j}\nonumber \\
 &-&\sum_\eta D^\eta\sum_{jk}b_{ik}^\eta b_{jk}^\eta\frac{\partial}{\partial m_j}\Bigg ) \rho\Bigg],
\end{eqnarray}
This equation describes the temporal evolution of the probability density $\rho(\boldsymbol{m},t)$ of finding a magnetic configuration with magnetization $\boldsymbol{m}$ at time $t$. In accordance with the ferromagnetic case we define $a_{\mathrm{A},i}(\boldsymbol{m},t)$ and $b_{\mathrm{A},ik}(\boldsymbol{m},t)$ for sublattice A as follows
\begin{eqnarray}
 \label{eq:A_LLB}
  a_{\mathrm{A},i}(\boldsymbol{m}_{\mathrm{A}},t)=&-& {\gamma'_{\mathrm{A}}\mu_0}\left( \boldsymbol{m}_{\mathrm{A}}\times \boldsymbol{H}_{\mathrm{eff}}\right) \nonumber \\
  &-&\frac{\alpha_{\mathrm{A}}^\perp  {\gamma'_{\mathrm{A}}\mu_0}}{m_{\mathrm{A}}^2} \left [ \boldsymbol{m}_{\mathrm{A}}\times \left ( \boldsymbol{m}_{\mathrm{A}}\times \boldsymbol{H}_{\mathrm{eff}} \right ) \right ]\nonumber \\
  &+&\frac{\alpha_{\mathrm{A}}^\parallel   {\gamma'_{\mathrm{A}}\mu_0}}{m_{\mathrm{A}}^2}\boldsymbol{m}_{\mathrm{A}}\left (\boldsymbol{m}_{\mathrm{A}}\cdot\boldsymbol{H}_{\mathrm{eff}}  \right ),
\end{eqnarray}
and
\begin{eqnarray}
 \label{eq:B_LLB}
 b^{\parallel}_{\mathrm{A},ik}(\boldsymbol{m}_{\mathrm{A}},t)&=&\delta_{ik}\nonumber\\
 b^{\perp}_{\mathrm{A},ik}(\boldsymbol{m}_{\mathrm{A}},t)&=&\alpha_{\mathrm{A}}^{\perp} \gamma'_{\mathrm{A}}\mu_0\left ( \delta_{ik}-\frac{m_{\mathrm{A},i}m_{\mathrm{A},k}}{m_{\mathrm{A}}^2} \right ).
\end{eqnarray}
Inserting Eqs.~\ref{eq:A_LLB} and \ref{eq:B_LLB} into Eq.~\ref{eq:FP} yields the FP equation for the sublattice
\begin{widetext}
\begin{eqnarray}
\label{eq:FP_LLB}
  \frac{\partial \rho_{\mathrm{A}}}{\partial t}=-\frac{\partial}{\partial \boldsymbol{m}_{\mathrm{A}}}\cdot\bigg \{ \bigg [ &-&{\gamma'_{\mathrm{A}}\mu_0}\left( \boldsymbol{m}_{\mathrm{A}}\times \boldsymbol{H}_{\mathrm{eff}}\right)-\frac{\alpha_{\mathrm{A}}^\perp  {\gamma'_{\mathrm{A}}\mu_0}}{m_{\mathrm{A}}^2}  \boldsymbol{m}_{\mathrm{A}}\times \left ( \boldsymbol{m}_{\mathrm{A}}\times \boldsymbol{H}_{\mathrm{eff}} \right )  \\
  &+&\frac{\alpha_{\mathrm{A}}^\parallel   {\gamma'_{\mathrm{A}}\mu_0}}{m_{\mathrm{A}}^2}\boldsymbol{m}_{\mathrm{A}}\left (\boldsymbol{m}_{\mathrm{A}}\cdot\boldsymbol{H}_{\mathrm{eff}}  \right ) 
  +\frac{D_{\mathrm{A}}^\perp(\alpha_{\mathrm{A}}^{\perp} \gamma'_{\mathrm{A}}\mu_0)^2}{m_{\mathrm{A}}^2} \boldsymbol{m}_{\mathrm{A}} \times \left ( \boldsymbol{m}_{\mathrm{A}} \times \frac{\partial }{\partial \boldsymbol{m}_{\mathrm{A}}}  \right )
  - D_{\mathrm{A}}^\parallel\frac{\partial}{\partial \boldsymbol{m}_{\mathrm{A}}}  \bigg ] \rho_{\mathrm{A}} \bigg \}.\nonumber
\end{eqnarray}
\end{widetext}
As already mentioned, the main objective is to determine the coefficients $D_\nu^\eta$, which are a measure for the magnitude of thermal fluctuations. To compute these coefficients we assume that in equilibrium the probability density of each sublattice magnetization follows a Boltzmann distribution per:
\begin{equation}
 \rho_{\mathrm{A}}=\rho_{\mathrm{A,}0}\exp(-E(\boldsymbol{m}_{\mathrm{A}})/k_{\mathrm{B}} T),
\end{equation}
\begin{equation}
 \label{eq:rho0}
 \frac{\partial \rho_{\mathrm{A}}}{\partial \boldsymbol{m}_{\mathrm{A}}}=\rho_{\mathrm{A}}\frac{\mu_0 M_{\mathrm{A,}0} V}{k_{\mathrm{B}} T}\boldsymbol{H}_{\mathrm{eff}}=\rho_{\mathrm{A}}\frac{\mu_0 n_\mathrm{at} x_{\mathrm{A}} \mu_{\mathrm{A}}V}{l_\mathrm{at}^3 k_{\mathrm{B}} T}\boldsymbol{H}_{\mathrm{eff}}.
\end{equation}
This equation holds for a discrete system with discretization volume $V$. In the last term of Eq.~\ref{eq:rho0} we identified the total magnetic moment of sublattice A with atomistic quantities. Here, $x_{\mathrm{A}}$ is the concentration of atoms A, $l_\mathrm{at}$ is the lattice constant and $n_\mathrm{at}$ is the number of atoms per unit cell. Using the expression of Eq.~\ref{eq:rho0} in the FP equation and demanding that $\partial \rho_{\mathrm{A}} / \partial t=0$ is valid in equilibrium, the diffusion constants of sublattice A can be computed to
\begin{equation}
\label{eq:D_perp}
 D_\mathrm{A}^\perp=\frac{\left (\alpha_\mathrm{A}^\perp-\alpha_\mathrm{A}^\parallel  \right ) l_\mathrm{at}^3 k_{\mathrm{B}}T}{(\alpha_\mathrm{A}^{\perp})^2 \gamma'_\mathrm{A} \mu^2_0 n_\mathrm{at} x_{\mathrm{A}} \mu_{\mathrm{A}}V}
\end{equation}
\begin{equation}
\label{eq:D_parallel}
  D_\mathrm{A}^\parallel=\frac{\alpha_{\mathrm{A}}^\parallel  \gamma'_\mathrm{A} l_\mathrm{at}^3 k_{\mathrm{B}}T}{ n_\mathrm{at} x_{\mathrm{A}} \mu_{\mathrm{A}}V}.
\end{equation}
Finally, the corresponding stochastic LLB equation for ferrimagnets can be obtained by using Eqs.~\ref{eq:A_LLB} and \ref{eq:B_LLB} together with Eqs.~\ref{eq:D_perp}, \ref{eq:D_parallel} and \ref{eq:fluct} in the Langevin equation (Eq.~\ref{eq:Langevin}) per
\begin{eqnarray}
 \label{eq:LLB_ferri_stoc}
 \frac{\partial \boldsymbol{m}_\mathrm{A}}{\partial t}=&-& \mu_0\gamma'_\mathrm{A} \left ( \boldsymbol{m}_\mathrm{A} \times \boldsymbol{H}_{\mathrm{eff,}\mathrm{A}} \right )\nonumber \\
 &+& \frac{\mu_0\gamma'_\mathrm{A} \alpha_\mathrm{A}^\parallel}{m_\mathrm{A}^2} \left ( \boldsymbol{m}_\mathrm{A} \cdot \boldsymbol{H}_{\mathrm{eff,}\mathrm{A}} \right )\boldsymbol{m}_\mathrm{A} + \boldsymbol{\xi}_{A}^\parallel  \\
 &-& \frac{\mu_0\gamma'_\mathrm{A} \alpha_\mathrm{A}^\perp}{m_\mathrm{A}^2}\left \{ \boldsymbol{m}_\mathrm{A} \times \left [ \boldsymbol{m}_\mathrm{A} \times \left (\boldsymbol{H}_{\mathrm{eff,}\mathrm{A}} + \boldsymbol{\xi}_{A}^\perp \right ) \right ] \right \}.\nonumber
\end{eqnarray}

\subsection{finite system susceptibilities}
\label{sec:sus}
To integrate the LLB equation detailed knowledge of the longitudinal susceptibilities $\tilde{\chi}_{\mathrm{A}}^\parallel$ and $\tilde{\chi}_{\mathrm{B}}^\parallel$ are required. In the original work of Atxitia~et~al.~\cite{atxitia_landau-lifshitz-bloch_2012} a mean field approach was derived
\begin{widetext}
\begin{equation}
 \label{eq:chi_para}
 \tilde{\chi}_{\mathrm{A,mean}}^\parallel= \frac{ \mu_{\mathrm{B}} \mathcal{L}'_{\mathrm{A}}(\zeta_{\mathrm{A}}) |J_{0\mathrm{,AB}}|  \mathcal{L}'_{\mathrm{B}}(\zeta_{\mathrm{B}}) + \mu_{\mathrm{A}}   \mathcal{L}'_{\mathrm{A}}(\zeta_{\mathrm{A}}) [k_{\mathrm{B}} T-J_{0\mathrm{,BB}}  \mathcal{L}'_{\mathrm{B}}(\zeta_{\mathrm{B}})] }{[k_{\mathrm{B}} T-J_{0\mathrm{,AA}}  \mathcal{L}'_{\mathrm{A}}(\zeta_{\mathrm{A}})][k_{\mathrm{B}} T-J_{0\mathrm{,BB}}  \mathcal{L}'_{\mathrm{B}}(\zeta_{\mathrm{B}})]-|J_{0\mathrm{,BA}}| \mathcal{L}'_{\mathrm{A}}(\zeta_{\mathrm{A}})|J_{0\mathrm{,AB}}| \mathcal{L}'_{\mathrm{B}}(\zeta_{\mathrm{B}})}.
\end{equation}
\end{widetext}
In this equation $\mathcal{L}_{\mathrm{A}}$ is the Langevin function with argument $\zeta_{\mathrm{A}}=(J_{0\mathrm{,AA}} m_\mathrm{A} + |J_{0\mathrm{,AB}}| m_\mathrm{B})/(k_\mathrm{B} T)$ and $\mathcal{L}'_{\mathrm{A}}$ is the corresponding derivative with respect to $\zeta_{\mathrm{A}}$. Equation~\ref{eq:chi_para} is, strictly speaking, correct only for infinite systems. Hence, to properly model a magnet with finite size other strategies are needed. This discrepancy was already extensively discussed in the case of pure ferromagnets~\cite{kazantseva_towards_2008,volger_llb,vogler_influence_2016}. Additionally, the importance of modeling the temperature dependence of the anisotropy field was shown. By means of the perpendicular susceptibility the anisotropy field in each sublattice can be defined as
\begin{equation}
 \label{eq:Hani}
 \boldsymbol{H}_{\mathrm{ani,A}}=\frac{1}{\tilde{\chi}_{\mathrm{A}}^\perp}(m_{x\mathrm{,A}}\boldsymbol{e}_x+m_{y\mathrm{,A}}\boldsymbol{e}_y).
\end{equation}
Here, the temperature dependence is included in $\tilde{\chi}_{\mathrm{A}}^\perp$. A benefit is that both parallel and perpendicular susceptibility can be computed from thermodynamics. Spin fluctuations at zero field along and perpendicular to the anisotropy axis can be used to derive an expression for the response function. How this is done for a ferromagnet is briefly reviewed in the following. The result will help to understand the response of susceptibilities of sublattices in a ferrimagnet.

The canonical partition function $Z$ of magnetization $\boldsymbol{M}_i$ in microstate $i$, which is subject to a field $\boldsymbol{B}$, can be expressed per:
\begin{equation}
\label{eq:Z}
 Z=\sum_i e^{-\beta\left( E_i-V \boldsymbol{M}_{i}\cdot \boldsymbol{B}\right)}.
\end{equation}
The expectation value of the magnetization can be written as
\begin{eqnarray}
\label{eq:meanM}
 \langle \boldsymbol{M} \rangle&=&\frac{1}{Z}\sum_i \boldsymbol{M}_{i}e^{-\beta\left( E_i-V \boldsymbol{M_{i}}\cdot \boldsymbol{B}\right)}\nonumber \\
 &=&\frac{1}{Z}\frac{1}{\beta V}\frac{\partial Z}{\partial \boldsymbol{B}}.
\end{eqnarray}
A similar expression for the expectation value of the squared magnetization can be easily found per
\begin{equation}
\label{eq:meanSquareMviaZ}
 \langle \boldsymbol{M}^2 \rangle=\frac{1}{Z}\frac{1}{\beta^2 V^2}\frac{\partial^2 Z}{\partial \boldsymbol{B}^2}.
\end{equation}
Based on the definition of the susceptibility
\begin{equation}
\label{eq:chiTildeDef}
  \boldsymbol{\chi}=\left( \frac{\partial \langle \boldsymbol{M} \rangle }{\partial \boldsymbol{H}}\right)_T=\mu_0\left( \frac{\partial \langle \boldsymbol{M} \rangle }{\partial \boldsymbol{B}}\right)_T,
\end{equation}
Eqs.~\ref{eq:meanM} and \ref{eq:meanSquareMviaZ} can be used to calculate $\chi$ per
\begin{equation}
\label{eq:chiFluc_vec}
  \boldsymbol{\chi}=\mu_0\beta V \left[\langle \boldsymbol{M}^2 \rangle-\langle \boldsymbol{M} \rangle^2\right].
\end{equation}
Obviously, the same expressions hold for the components of the susceptibility 
\begin{equation}
\label{eq:chiFluc}
  \chi^\eta=\mu_0\beta V \left[\langle M_\eta^2 \rangle-\langle M_\eta \rangle^2\right].
\end{equation}
We now assume that the ferromagnet is split into two sublattices with concentrations $x_{\mathrm{A}}$ and $x_{\mathrm{B}}$, with $x_{\mathrm{A}}+x_{\mathrm{B}}=1$. Hence, the partition function can be written as
\begin{equation}
\label{eq:Zferri}
 Z=\sum_i e^{-\beta\left[ E_i-(x_{\mathrm{A}}+x_{\mathrm{B}})V\boldsymbol{M}_{i}\cdot \boldsymbol{B}\right]}.
\end{equation}
The same procedure as shown above can now be applied to obtain the susceptibility
\begin{equation}
\label{eq:chiFlucFerri}
  \chi^\eta=\mu_0\beta (x_{\mathrm{A}}+x_{\mathrm{B}})V \left[\langle M_\eta^2 \rangle-\langle M_\eta \rangle^2\right].
\end{equation}
Obviously, $\chi^\eta$ can be divided into two expressions for the corresponding sublattices. Without loss of generality we further analyze the longitudinal susceptibility of sublattice A
\begin{eqnarray}
\label{eq:chiFlucFerriA}
  \chi^\parallel_{\mathrm{A}}&=&\mu_0 \beta x_{\mathrm{A}}V \left[\langle M_z^2 \rangle-\langle M_z \rangle^2\right]\nonumber \\
  &=& \frac{\mu_0 \beta}{x_{\mathrm{A}}V}  \left[\langle (x_{\mathrm{A}}VM_z)^2 \rangle-\langle x_{\mathrm{A}}VM_z \rangle^2\right].
\end{eqnarray}
The expression $x_{\mathrm{A}}VM_z=\sum_i^{N_\mathrm{A}} \boldsymbol{e}_z \cdot \boldsymbol{\mu}_i$ can be identified with the total magnetic moment of sublattice A in z direction resulting in
\begin{equation}
\label{eq:chiFlucFerriA_2}
  \chi^\parallel_{\mathrm{A}}=\frac{\mu_0 \beta \left(N_\mathrm{A}\mu_\mathrm{A} \right)^2}{x_{\mathrm{A}}V} \left[ \left \langle m_{\mathrm{A}}^2 \right \rangle- \left\langle m_{\mathrm{A}} \right \rangle^2\right],
\end{equation}
with the normalized magnetization of the sublattice
\begin{equation}
 m_{\mathrm{A}}=\frac{\sum_i^{N_\mathrm{A}}\boldsymbol{e}_z \cdot \boldsymbol{\mu}_i}{N_\mathrm{A}\mu_\mathrm{A}}.
\end{equation}
In Eq.~\ref{eq:lambda} we are interested in the quantity $\tilde{\chi}^\parallel_{\mathrm{A}}=\chi^\parallel_{\mathrm{A}}/(\mu_0 M_{\mathrm{A,}0})$. Hence, the final expression takes the following form
\begin{eqnarray}
\label{eq:chiFlucFerriA_final}
  \tilde{\chi}_{\mathrm{A}}^\parallel&=&\frac{\beta \left(N_\mathrm{A}\mu_\mathrm{A} \right)^2}{x_{\mathrm{A}} V} \frac{ l_\mathrm{at}^3}{n_\mathrm{at} x_{\mathrm{A}} \mu_{\mathrm{A}}} \left[ \left \langle m_{\mathrm{A}}^2 \right \rangle- \left\langle m_{\mathrm{A}} \right \rangle^2\right]\nonumber \\
  &=&\frac{N_\mathrm{A}\mu_\mathrm{A}}{k_{\mathrm{B}} T} \frac{ 1}{x_{\mathrm{A}}} \left[ \left \langle m_{\mathrm{A}}^2 \right \rangle- \left\langle m_{\mathrm{A}} \right \rangle^2\right].
\end{eqnarray}
In contrast to $\tilde{\chi}^\parallel$ of the whole system a factor $x_{\mathrm{A}}^{-1}$ appears in the susceptibility of the ferromagnetic sublattice $\tilde{\chi}_{\mathrm{A}}^\parallel$, which is an important but non-obvious result. Since a ferrimagnet consists of two ferromagnetic sublattices we need Eq.~\ref{eq:chiFlucFerriA_final} to correctly extract the sublattice susceptibilities from spin fluctuations.

\subsection{material function scaling}
\label{sec:material_function_scaling}
The susceptibilities obtained must be adjusted before being entered in the LLB equation via the effective field. Typically, functions from a mean field model are fitted for this purpose. To show how this procedure works for ferrimagnets we would like to rely on an example. For better comparability with Ref.~\cite{atxitia_landau-lifshitz-bloch_2012} we use a cylindrical nanoparticle consisting of GdFeCo as sample system. The geometry and the material parameters of the particle are shown in Tab.~\ref{tab:material}.
\begin{table}[h!]
  \centering
  \vspace{0.5cm}
  \begin{tabular}{c c c}
    \toprule
    \toprule
      & A (FeCo) & B (Gd) \\
    \midrule
    $d$\,[J] & $8.07251\mathrm{x}10^{-24}$ & $8.07251\mathrm{x}10^{-24}$ \\
    $\mu$\,[$\mu_{\mathrm{Bohr}}$] & 2.217 & 7.63 \\
    $x$& 0.7 & 0.3 \\
    $J_{\kappa-\kappa}$\,[J] & $4.5\mathrm{x}10^{-21}$ & $1.26\mathrm{x}10^{-21}$ \\
    $J_{\kappa-\nu}$\,[J] & \multicolumn{2}{c}{$-1.09\mathrm{x}10^{-21}$} \\
    \midrule
    $n_{\mathrm{at}}$ & \multicolumn{2}{c}{4} \\
    $r$\,[nm] & \multicolumn{2}{c}{5.0} \\
    $h$\,[nm] & \multicolumn{2}{c}{10.0} \\
    \midrule
    $T_{\mathrm{C}}$\,[K] & \multicolumn{2}{c}{697} \\
    $T_{\mathrm{comp}}$\,[K] & \multicolumn{2}{c}{313} \\
    \bottomrule
    \bottomrule
  \end{tabular}
  \caption{\small Geometry and material parameters of both sublattices A and B in GdFeCo (taken from Ref.~\cite{atxitia_landau-lifshitz-bloch_2012}). $d$ is the anisotropy energy per atom, $\mu$ is the magnetic moment in units of Bohr magnetons, $x$ is the concentration, $J_{\kappa-\kappa}$ denotes the exchange energy per atom link between equal atoms, $J_{\kappa-\nu}$ denotes the exchange energy per atom link between different atoms, $n_{\mathrm{at}}$ is the number of atoms per unit cell, $l_{\mathrm{at}}$ is the lattice parameter and $r$ and $h$ are the radius and the height of the particle. Curie temperature and compensation point are denoted with $T_{\mathrm{C}}$ and $T_{\mathrm{comp}}$, respectively.}
  \label{tab:material}
\end{table}

In order to be able to quantitatively and qualitatively validate the results of the proposed coarse grained LLB model we use a finite difference model with atomistic discretization as reference. The magnetization dynamics of this reference model are assumed to be correct in a sense that we aim to reproduce them with the presented coarse grained ferriLLB model. We use the atomistic code VAMPIRE~\cite{evans_atomistic_2014} solving the stochastic Landau-Lifshitz-Gilbert equation for each spin. VAMPIRE is also used to compute the temperature dependent average magnetization and the temperature dependent spin fluctuations in order to determine the needed input functions (magnetization and susceptibilities) for the integration of the ferriLLB equation. For this purpose system trajectories with $10^7$ time steps (after $2\mathrm{x}10^4$ equilibration steps) with an integration time step of $10^{-15}$\,s for each temperature value in the range of $0-950$\,K are simulated by means of a stochastic Heun integration schema. 
\begin{figure}[!h]
\includegraphics{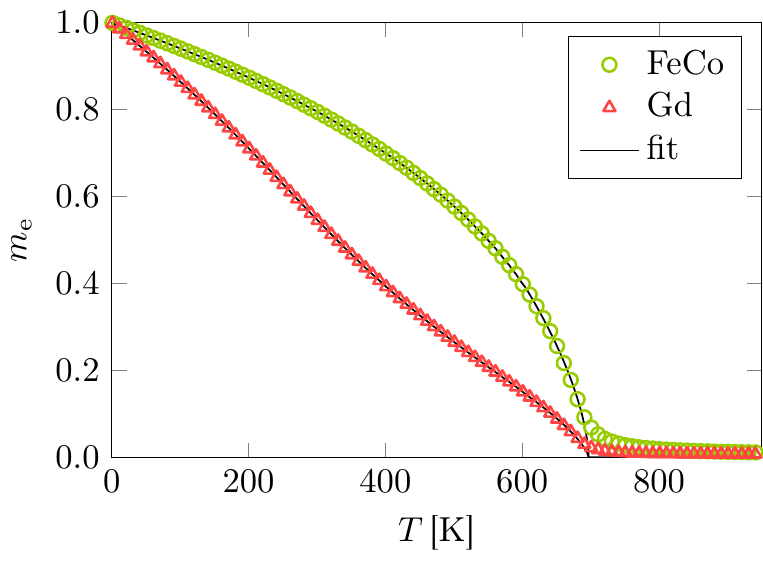}
  \caption{\small (color online) Zero field equilibrium magnetization $m_{\mathrm{e}}$ versus temperature, computed with an atomistic model of GdFeCo (parameters are given in Tab.~\ref{tab:material}). The black solid lines show fits, representing an infinite system.}
  \label{fig:me_fit}
\end{figure}
Firgure~\ref{fig:me_fit} displays the resulting equilibrium magnetization at zero field for both sublattices. To use the data in Eq.~\ref{eq:LLB_Hpara} we first fit the FeCo curve $m_{\mathrm{e,A}}$ with the mean field expression
\begin{equation}
 m_{\mathrm{e}}(T)=c_1\left ( 1- \frac{T}{c_2} \right)^{c_3},
\end{equation}
with fit parameters $c_1, c_2$ and $c_3$. Here, the Curie temperature $T_{\mathrm{C}}=c_2=697$\,K of the ferrimagnet is determined. In the fit procedure of the second sublattice this Curie temperature is fixed and just the other two parameters are adjusted. The resulting fit functions are plotted in Fig.~\ref{fig:me_fit} with black solid lines.

\begin{figure}[!h]
\includegraphics{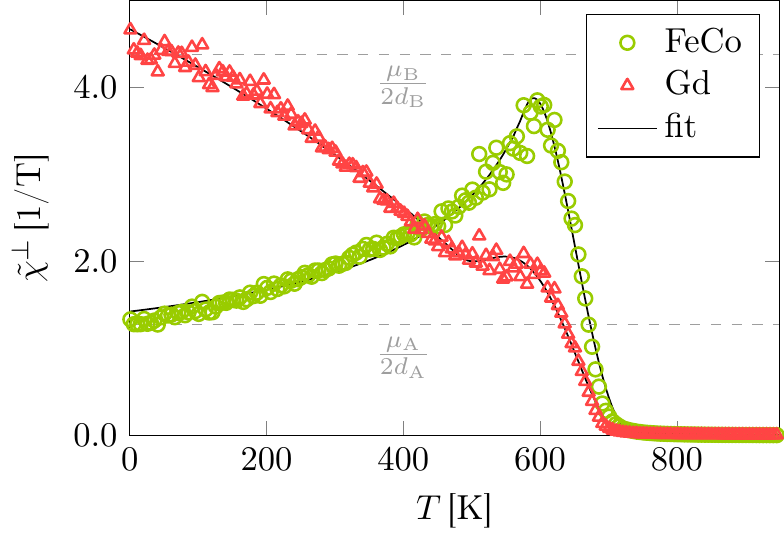}
  \caption{\small (color online) Perpendicular susceptibility $\tilde{\chi}^\perp$, computed with an atomistic model of GdFeCo (parameters are given in Tab.~\ref{tab:material}) from magnetization fluctuations. The black solid lines show fits, representing an infinite system.}
  \label{fig:cho_normal_fit}
\end{figure}

The same trajectories from which the equilibrium magnetizations were determined can also be used to calculate the fluctuations of the magnetization parallel and perpendicular to the anisotropy axis by means of Eq.~\ref{eq:chiFlucFerriA_final}. $\tilde{\chi}^\perp$ for both sublattices is shown in Fig.~\ref{fig:cho_normal_fit}. 
With the expression (Eq.~\ref{eq:chiFlucFerriA_final}) derived in Sec.~\ref{sec:sus} the susceptibilities agree well with the inverse anisotropy field at zero temperature, which is also displayed as dashed line in Fig.~\ref{fig:cho_normal_fit} for both sublattices. Note, that the susceptibilities change considerably with temperature. This fact suggests that it is very important to correctly model the temperature dependence of $\tilde{\chi}^\perp$ and not only to use the zero temperature value for the whole temperature range. A detailed comparison will be presented in Sec.~\ref{sec:Results}. To extract the susceptibilities for the usage in Eq.~\ref{eq:LLB_Hpara} we use the same fitting procedure as proposed in Ref.~\cite{volger_llb} for ferromagnets per
\begin{equation}
 \widetilde{\chi}_\perp(T)=\begin{cases}
 c_4 m_{\mathrm{e}}^{c_5} & T<<T_\mathrm{C} \\
 \frac{c_6}{T-T_\mathrm{C}} & T>T_\mathrm{C}
\end{cases}.
\end{equation}

\begin{figure}[!h]
\includegraphics{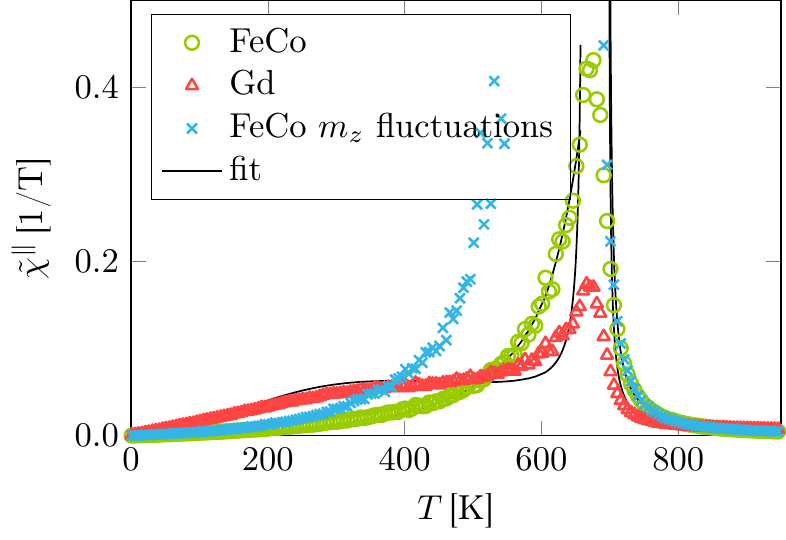}
  \caption{\small (color online) Longitudinal susceptibility $\tilde{\chi}^\parallel$, computed with an atomistic model of GdFeCo (parameters are given in Tab.~\ref{tab:material}) from magnetization fluctuations. The black solid lines show fits, representing an infinite system.}
  \label{fig:cho_para_fit}
\end{figure}
Parallel susceptibilities are presented in Fig.~\ref{fig:cho_para_fit}. As suggested and explained in detail in Ref.~\cite{volger_llb} we use the fluctuations of the magnitude of the magnetization to determine the parallel susceptibilities. Since both sublattices are soft magnetic the fluctuations of $m_z$ are too noisy near the Curie temperature to be able to extract the true parallel susceptibilities from them, as pointed out in Fig.~\ref{fig:cho_para_fit} for the FeCo sublattice. As fit function for $\tilde{\chi}^\parallel$ we use the mean field expression of Eq.~\ref{eq:chi_para} with two fit parameters $c_7$ and $c_8$ as follows
\begin{equation}
 \label{eq:chi_para_fit_func}
  \tilde{\chi}^\parallel(T)=c_7\tilde{\chi}_{\mathrm{mean}}^\parallel(c_8J_{0\mathrm{,AA}},c_8J_{0\mathrm{,BB}},c_8J_{0\mathrm{,AB}},c_8J_{0\mathrm{,BA}},T).
\end{equation}
This means that each exchange energy appearing in Eq.~\ref{eq:chi_para} is scaled by the fit parameter $c_8$, which is equivalent to a scaling of the Curie temperature. To understand this behavior the denominator of Eq.~\ref{eq:chi_para} can be analyzed. Since the susceptibility diverges at $T_{\mathrm{C}}$ the denominator becomes zero. With this condition the mean field Curie temperature can be determined to
\begin{eqnarray}
  \label{eq:Tc}
 T_{\mathrm{C,mean}}=\frac{1}{6k_{\mathrm{B}}}&\Big( &\sqrt{(J_{0\mathrm{,AA}}-J_{0\mathrm{,BB}})^2-4J_{0\mathrm{,AB}}J_{0\mathrm{,BA}}}\nonumber \\
 &+& J_{0\mathrm{,AA}}+J_{0\mathrm{,BB}}  \Big ).
\end{eqnarray}
From Eq.~\ref{eq:Tc} it becomes clear that a scaling of all exchange energies is equivalent to a scaling of $T_{\mathrm{C}}$. But, just shifting the Curie temperature is not enough to adapt the susceptibilities to the correct finite size behavior. Scaling of the whole susceptibility function is additionally required via fit parameter $c_7$ of Eq.~\ref{eq:chi_para_fit_func}.

Another issue that needs to be clarified is the meaning of the expression $\Lambda_1$ at $T_{\mathrm{C}}$ in Eq.~\ref{eq:lambda}. Since both susceptibilities diverge, the limit of the quotient $\tilde{\chi}_{\mathrm{B}}^\parallel/\tilde{\chi}_{\mathrm{A}}^\parallel$ must be determined. Nieves~et~al.~\cite{nieves_classical_2015} derived a compact form of $\Lambda_1$ at $T_{\mathrm{C}}$ per
\begin{equation}
 \label{eq:lambda_nieves}
 \Lambda_1=\frac{3k_{\mathrm{B}}T_{\mathrm{C}}-c_8J_{0\mathrm{,AA}}}{\mu_{\mathrm{A}}}.
\end{equation}
Note, in this equation the scaling parameter $c_8$ is again needed to ensure that the exchange energy $J_{0\mathrm{,AA}}$ yields the finite size Curie temperature. Near $T_{\mathrm{C}}$ Eq.~\ref{eq:lambda_nieves} (instead of Eq.~\ref{eq:lambda}) is used in Eq.~\ref{eq:LLB_Hpara} in the coarse grained ferriLLB model.

\section{Results}
\label{sec:Results}
In order to confirm the validity of the proposed coarse grained model numerical tests for the presented GdFeCo system (see Tab.~\ref{tab:material}) are performed in the following. First, the dynamics of single magnetization trajectories under the influence of heat and magnetic field are compared with corresponding trajectories computed with the atomistic code VAMPIRE. 
\begin{figure}
    \centering
    \includegraphics{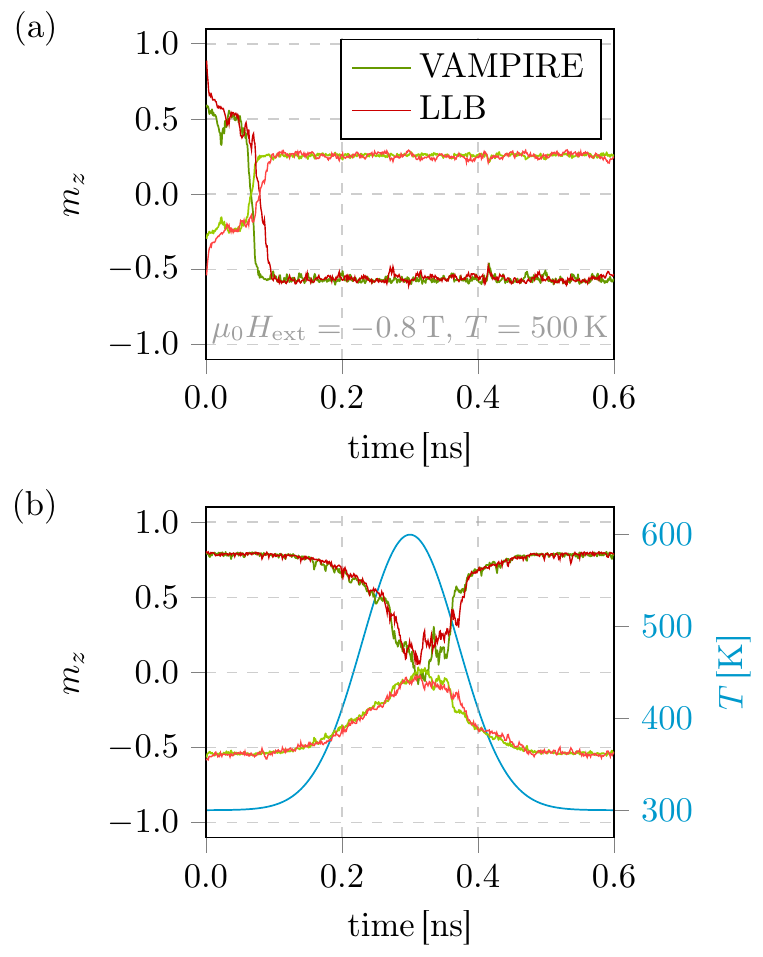}
    \caption{\small (color online) Temporal evolution of the z component of the normalized magnetization of both sublattices of GdFeCo computed with the proposed coarse grained ferriLLB model and the atomistic code VAMPIRE. (a) A constant magnetic field with $\mu_0 H_{\mathrm{ext}}=-0.8$\,T and an angle of 6\,$^\circ$ with the z direction is applied. (b) A Gaussian shaped heat pulse is applied (blue solid line, right y axis).}
    \label{fig:trajectories}
\end{figure}
In Fig.~\ref{fig:trajectories}a) a constant temperature of 500\,K and a constant magnetic field of $-0.8$\,T are applied to the ferrimagnet. Field and easy axis of the grain (along z direction) enclose an angle of 6\,$^\circ$. 500\,K is well above the compensation point and the ferrimagnet is FeCo dominated. The simulations are started with an initial magnetization of the FeCo sublattice in the positive z direction and the Gd sublattice magnetization pointing in the negative z direction. Unless otherwise stated, this initial configuration is used for all subsequent simulations. Figure~\ref{fig:trajectories}a) illustrates that the temporal evolution of $m_z$ of both sublattices obtained by the proposed coarse grained model agrees very well with the resulting VAMPIRE trajectories. 

In a second test we investigate the magnetization dynamics under a heat pulse, without an external field. A Gaussian shaped heat pulse is used
\begin{equation}
 \label{eq:Gaussian}
 T(t)=T_{\mathrm{min}}+(T_{\mathrm{max}}-T_{\mathrm{min}})e^{\frac{(t-t_0)^2}{\tau^2}},
\end{equation}
with $T_{\mathrm{min}}=300$\,K, $T_{\mathrm{max}}=600$\,K, $t_0=0.3$\,ns and $\tau=0.1$\,ns. The temperature pulse starts slightly below the compensation point and heats the ferrimagnet near $T_{C}$, before the system cools down again. Temperature pulse and $m_z$ of both sublattices are shown in Fig.~\ref{fig:trajectories}b). The results of our coarse grained model and VAMPIRE again agree perfectly.

In a next step hysteresis loops at constant temperatures are compared. We analyze easy axis loops and loops with a field angle of 45\,$^\circ$ with respect to the easy axis of the ferrimagnet. The loops start with a saturating field with a magnitude of 3\,T, which is decreased with a rate of 1\,T per nanosecond until $-3$\,T is reached. After that the field is again increased to 3\,T. The choice of the fast field rate results from the high computational effort of atomistic simulations. All loops are calculated at two different temperatures, 100\,K and 500\,K.
\begin{figure}
    \centering
    \begin{adjustwidth}{-0.3cm}{-0.3cm}
    \includegraphics{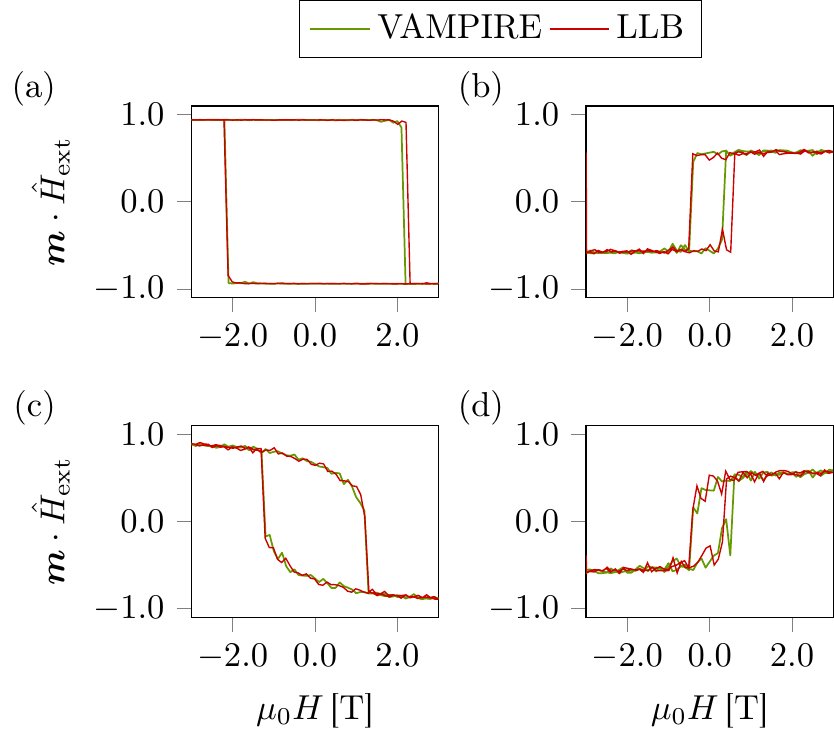}
    \caption{\small (color online) Hysteresis loops of GdFeCo with a field rate of 1\,T/ns calculated with the proposed coarse grained ferriLLB model and the atomistic code VAMPIRE. (a) Easy axis loop at a constant temperature of (a) 100\,K and (b) 500\,K. Hysteresis loop with the applied field tilted 45\,$^\circ$ against the z direction at a constant temperature of (c) 100\,K and (d) 500\,K.}
    \label{fig:hyst}
  \end{adjustwidth}
\end{figure}
Figure~\ref{fig:hyst} displays the calculated hysteresis loops of the total normalized magnetization of the ferrimagnet for the four cases. Again, the coarse grained ferriLLB model is in good agreement with atomistic VAMPIRE simulations.

In a last validation step switching probabilities of GdFeCo under the influence of various Gaussian heat pulses and a constant external field are analyzed. Again, a field with a magnitude of -0.8\,T and a field angle of 6\,$^\circ$ with the z direction tries to align the total magnetization of the ferrimagnet along the negative z direction. Additionally, a heat pulse, according to Eq.~\ref{eq:Gaussian}, with $T_\mathrm{min}=300\,K$ and various $T_\mathrm{max}$ is applied to the ferrimagnetic particle. For each $T_\mathrm{max}$, from 300\,K to 680\,K with $\Delta T_\mathrm{max}=20$\,K, 128 trajectories are computed. The switching probability then corresponds to the proportion of successfully aligned particles compared to the total number of all started simulations.
\begin{figure}[!h]
\includegraphics{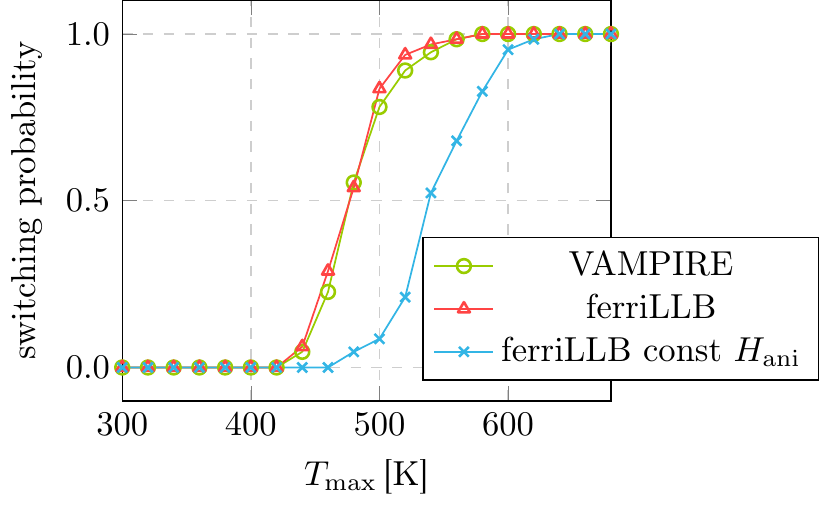}
  \caption{\small (color online) Switching probabilities of a GdFeCo particle computed from 128 switching trajectories at each $T_{\mathrm{max}}$. In each simulation a constant field with $\mu_0 H_{\mathrm{ext}}=-0.8$\,T and a Gaussian shaped heat pulse according to Eq.~\ref{eq:Gaussian} with $T_{\mathrm{min}}=300$\,K and $\tau=0.1$\,ns are applied.}
  \label{fig:porb_ferrriLLB}
\end{figure}
The comparison of the switching probabilities obtained by the coarse grained ferriLLB model and VAMPIRE simulations in Fig.~\ref{fig:porb_ferrriLLB} confirms the desired perfect agreement of the ferriLLB model. To check the influence of the temperature dependence of the perpendicular susceptibility in the ferriLLB model, which was introduced in Sec.~\ref{sec:material_function_scaling}, the probabilities are recomputed with the same setup, with the only difference that a constant anisotropy field $H_{\mathrm{ani,A}}=2d_{\mathrm{A}}/\mu_{\mathrm{A}}$, is used. The resulting probabilities, as illustrated in Fig.~\ref{fig:porb_ferrriLLB}, show a completely different behavior. This fact strengthens the conclusion that it is important to consider the temperature dependence of the anisotropy field in the coarse grained ferriLLB model.

\subsection{Equivalence of the ferromagnetic LLB equation}
\label{sec:Discussion}
As already mentioned the ferriLLB equation for each sublattice (Eq.~\ref{eq:LLB_ferri_stoc}) has the same form as the ferroLLB equation (Eq.~\ref{eq:ferroLLB}). At first glance they differ only in the effective field. In this section we derive an expression for the intergrain exchange field for two ferromagnetic sublattices, which couples their ferroLLB equations. Further we show that the resulting effective field is very similar to Eq.~\ref{eq:LLB_Hpara} and that using the same input functions (zero field magnetization and susceptibilities) the magnetization dynamics of a ferrimagnet can be computed equally with both, the ferriLLB equation as well as the ferroLLB together with the derived intergrain exchange field. 

To compute the intergrain exchange field between two ferromagnetic sublattices we refer to Ref.~\cite{volger_llb}, where the desired intergrain exchange field was deduced by determining the number of interacting spins between two coupled ferromagnetic layers on the boundary surface. Here, we follow the same strategy by computing the mean number of interacting spins between the two ferromagnetic sublattices. The Heisenberg Hamiltonian serves as a starting point
\begin{equation}
 \mathcal{H}=-\frac{1}{2}\sum_{\mathrm{nn}}J_{kl}\boldsymbol{s}_k \cdot \boldsymbol{s}_l,
\end{equation}
where $l$ and $k$ are lattice sites and $\boldsymbol{s}_k=\boldsymbol{\mu}_k/\mu_k$ denotes the unit vector of the magnetic moment on lattice site $k$. To obtain the intergrain exchange energy the sum just goes over all neighboring lattice sites nn which are occupied with atoms of different sublattices. We assume that the exchange integrals $J_{kl}$ are independent from the lattice site
\begin{equation}
 \mathcal{H}=-\frac{1}{2}J\sum_{\mathrm{nn}}\boldsymbol{s}_k \cdot \boldsymbol{s}_l.
\end{equation}
If $z$ is the number of nearest neighbors in the lattice each atom A has on average $zx_{\mathrm{B}}$ neighbors B and each atom B has on average $zx_{\mathrm{A}}$ neighbors A. Hence, the sum can be rewritten over all interacting pairs
\begin{equation}
 \mathcal{H}=-\frac{1}{2}Jzx_{\mathrm{B}}\sum_{\mathrm{i=1}}^{N_{\mathrm{A}}}\boldsymbol{s}_{i{\mathrm{,A}}} \cdot \boldsymbol{s}_{i{\mathrm{,B}}}-\frac{1}{2}Jzx_{\mathrm{A}}\sum_{\mathrm{j=1}}^{N_{\mathrm{B}}}\boldsymbol{s}_{j{\mathrm{,B}}} \cdot \boldsymbol{s}_{j{\mathrm{,A}}}.
\end{equation}
As explained in Ref.~\cite{volger_llb} the next step is the transition from the atomistic to the LLB description, where each sublattice is represented by one magnetization vector
\begin{equation}
 \label{eq:Hamiltonian1}
 \mathcal{H}=-\frac{1}{2}Jz(N_{\mathrm{A}}x_{\mathrm{B}}+N_{\mathrm{B}}x_{\mathrm{A}})\frac{\boldsymbol{m}_{\mathrm{A}}}{m_{\mathrm{A}}}\cdot\frac{\boldsymbol{m}_{\mathrm{B}}}{m_{\mathrm{B}}}.
\end{equation}
Since, $N_\nu=x_\nu V n_{\mathrm{at}}/l_{\mathrm{at}}^3$, Eq.~\ref{eq:Hamiltonian1} becomes
\begin{equation}
 \label{eq:Hamiltonian}
 \mathcal{H}=-Jzn_{\mathrm{at}}x_{\mathrm{A}}x_{\mathrm{B}}\frac{V}{l_{\mathrm{at}}^3}\left(\frac{\boldsymbol{m}_{\mathrm{A}}}{m_{\mathrm{A}}}\cdot\frac{\boldsymbol{m}_{\mathrm{B}}}{m_{\mathrm{B}}} \right).
\end{equation}
Finally, the intergrain exchange field of sublattice A is computed per
\begin{equation}
\label{eq:exchange_field_def}
 \mu_0\boldsymbol{H}_{\mathrm{iex,A}} = -\frac{1}{VM_{\mathrm{A},0}}\frac{\partial}{\partial \boldsymbol{m}_{\mathrm{A}}}\mathcal{H}.
\end{equation}
Keeping in mind that the absolute value $m_{\mathrm{A}}$ can be written as $\sqrt{\boldsymbol{m}_{\mathrm{A}}\cdot\boldsymbol{m}_{\mathrm{A}}}$ and with the definitions $M_{\mathrm{A},0}=n_{\mathrm{at}}x_{\mathrm{A}}\mu_{\mathrm{A}}/l_{\mathrm{at}}^3$ and $J_{0,\mathrm{A}\mathrm{B}}=zx_\mathrm{B} J$, the intergrain exchange field yields
\begin{eqnarray}
\label{eq:exchange_field}
 \mu_0\boldsymbol{H}_{\mathrm{iex,A}} =&& \frac{J_{0,\mathrm{A}\mathrm{B}}}{\mu_{\mathrm{A}}}\frac{\sqrt{m_{\mathrm{e,A}}^\alpha(T) m_{\mathrm{e,B}}^\beta(T)}}{m_{\mathrm{A}}m_{\mathrm{B}}} \nonumber \\
 &\cdot& \left ( \boldsymbol{m}_{\mathrm{B}} - \frac{\boldsymbol{m}_{\mathrm{A}}\cdot \boldsymbol{m}_{\mathrm{B}}}{m_{\mathrm{A}}^2} \boldsymbol{m}_{\mathrm{A}}\right ).
\end{eqnarray}
The factor $\sqrt{m_{\mathrm{e,A}}^\alpha(T) m_{\mathrm{e,B}}^\beta(T)}$ was introduced in Ref.~\cite{volger_llb} to account for the temperature dependence of the exchange constant. $\alpha$ and $\beta$ are power law exponents describing the temperature dependence of the bulk exchange constant in the sublattices. For a generic soft magnetic ferromagnet the exponent is $1.66$.

The intergrain exchange field of Eq.~\ref{eq:exchange_field} together with the ferroLLB equation in each sublattice is now used to compute the same switching probabilities as in Sec.~\ref{sec:Results}. Note, we use the temperature dependent functions $m_{\mathrm{e}}(T)$, $\tilde{\chi}_\parallel(T)$ and $\tilde{\chi}_\perp(T)$ determined for the sublattices of the ferrimagnet in Sec.~\ref{sec:material_function_scaling}, which were also used in the ferriLLB equation. 
\begin{figure}[!h]
\includegraphics{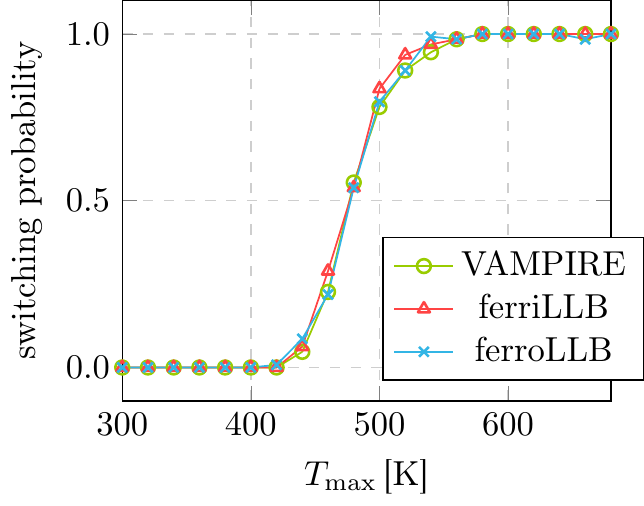}
  \caption{\small (color online) Switching probabilities of a GdFeCo particle computed for the same setup as shown in Fig.~\ref{fig:porb_ferrriLLB}. Here the results of the atomistic code VAMPIRE, the proposed coarse grained ferriLLB model are again illustrated. Additionally, each sublattice is computed with the coarse grained ferroLLB model (see Appendix~\ref{sec:appendix}) and coupled via the derived intergrain exchange field of Eq.~\ref{eq:exchange_field}.}
  \label{fig:porb_ferrroLLB}
\end{figure}
The resulting switching probabilities in Fig.~\ref{fig:porb_ferrroLLB} display that using the proper intergrain exchange field the ferroLLB equation yields the same agreement with VAMPIRE simulations as the more complex ferriLLB equation. This agreement might be a bit surprising at first glance, but if the effective fields of ferroLLB and ferriLLB equation are examined more closely the similarities become obvious.

First of all, in the proposed form the anisotropy fields $\boldsymbol{H}_{\mathrm{ani,A}}$ (compare Eq.~\ref{eq:Hani} and Eq.~\ref{eq:HaniFerro}) are identical. Further, the third term in Eq.~\ref{eq:LLB_Hpara} is a vector normal to $\boldsymbol{m}_{\mathrm{A}}$, and thus is only not vanishing if it enters into the terms of the ferriLLB that change the direction of the magnetization. If the identity of the double cross product is used we obtain
\begin{eqnarray}
\label{eq:doubleCross}
 &&-\frac{J_{0,\mathrm{A}\mathrm{B}}}{\mu_\mathrm{A} m_\mathrm{A}^2}\left[\boldsymbol{m}_\mathrm{A}\times\left( \boldsymbol{m}_\mathrm{A} \times \boldsymbol{m}_\mathrm{B}\right)\right]\nonumber \\
 &=&\frac{J_{0,\mathrm{A}\mathrm{B}}}{\mu_\mathrm{A}} \boldsymbol{m}_\mathrm{B}-\frac{J_{0,\mathrm{A}\mathrm{B}}}{\mu_\mathrm{A} m_\mathrm{A}^2}\left(\boldsymbol{m}_\mathrm{A}\cdot\boldsymbol{m}_\mathrm{B}\right)\boldsymbol{m}_\mathrm{A}.
\end{eqnarray}
The second term in Eq.~\ref{eq:doubleCross} does not influence the magnetization dynamics. If the ferrimagnet is near equilibrium, which is a good assumption for the majority of the simulation time due to the rapid longitudinal relaxation of the LLB equation, the first term in Eq.~\ref{eq:exchange_field} corresponds to the remaining term of Eq.~\ref{eq:doubleCross}.

The effective field of both formulations has a term which quickly relaxes the magnitude of the magnetization to its equilibrium value. In the ferriLLB equation the corresponding field term consists of two contributions as can be seen in Eq.~\ref{eq:lambda}. Only the first term has its counterpart in the ferroLLB equation (Eq.~\ref{eq:blochField}). Nevertheless, the second term 
\begin{equation}
 \frac{\tilde{\chi}_\mathrm{B}^\parallel}{2 \tilde{\chi}_\mathrm{A}^\parallel} \frac{|J_{0,\mathrm{AB}}|}{\mu_\mathrm{A}}
\end{equation}
is only dominating very close to $T_{\mathrm{C}}$, where the susceptibilities diverge, while the quotient $\tilde{\chi}_\mathrm{B}^\parallel/\tilde{\chi}_\mathrm{A}^\parallel$ remains finite. In this small range the ferriLLB equation shows a faster relaxation of the sublattice magnetizations towards its equilibrium value, compared to the ferroLLB equation. Obviously, this faster relaxation has not a large influence up to the simulated temperatures.

Additionally, the fourth term of Eq.~\ref{eq:LLB_Hpara} controls the angle between the magnetization of both sublattices. Under the assumption that the magnetizations are near equilibrium we can expand $(\boldsymbol{m}_\mathrm{A} \cdot \boldsymbol{m}_\mathrm{B})$ around $(\boldsymbol{m}_{\mathrm{e},\mathrm{A}} \cdot \boldsymbol{m}_{\mathrm{e},\mathrm{B}})$, yielding
\begin{eqnarray}
 &&1-\left(\frac{\boldsymbol{m}_\mathrm{A} \cdot \boldsymbol{m}_\mathrm{B}}{\boldsymbol{m}_{\mathrm{e},\mathrm{A}} \cdot \boldsymbol{m}_{\mathrm{e},\mathrm{B}}}\right)^2\frac{m_{\mathrm{e},\mathrm{A}}^2}{m_\mathrm{A}^2}\nonumber\\
 &\approx&-2 \left|\frac{\boldsymbol{m}_\mathrm{A} \cdot \boldsymbol{m}_\mathrm{B}}{\boldsymbol{m}_{\mathrm{e},\mathrm{A}} \cdot \boldsymbol{m}_{\mathrm{e},\mathrm{B}}}\right|\frac{m_{\mathrm{e},\mathrm{A}}^2}{m_\mathrm{A}^2}+1+\frac{m_\mathrm{B}^2}{m_{\mathrm{e},\mathrm{B}}^2}.
\end{eqnarray}
Together with the prefactor the first term of this equation becomes
\begin{equation}
  -\frac{|J_{0,\mathrm{AB}}|}{\mu_\mathrm{A}}\frac{\left|\boldsymbol{m}_\mathrm{A} \cdot \boldsymbol{m}_\mathrm{B}\right|}{m_\mathrm{A}^2}\boldsymbol{m}_\mathrm{A}.
\end{equation}
Near equilibrium this expression corresponds well to the second term of the derived intergrain exchange field in Eq.~\ref{eq:exchange_field}.

In a nutshell, we have shown that almost every effective field term of the ferriLLB equation has its counterpart in the ferroLLB equation if the derived intergrain exchange field of Eq.~\ref{eq:exchange_field} is used to couple the ferromagnetic sublattices. As a consequence the good agreement of simulated switching probabilities with both equations in Fig.~\ref{fig:porb_ferrroLLB} can be well understood.

\section{Conclusion}
\label{sec:conclusion}
In this work we developed a coarse grained model of disordered ferrimagnets based on the ferrimagnetic Landau-Lifshitz-Bloch (ferriLLB) equation~\cite{atxitia_landau-lifshitz-bloch_2012}. In a first step, stochastic fields were incorporated into the ferriLLB equation in order to account for thermal fluctuations of individual system trajectories. In a second step, an expression for the susceptibilities of finite sized ferrimagnets was derived from thermodynamics. As with the LLB equation of ferromagnets (ferroLLB), modeling these temperature-dependent material functions, including the zero field equilibrium magnetization, is the key to accurately describing the magnetization dynamics of ferrimagnets with high computational efficiency. We have shown that the presented coarse grained model agrees well with atomistic simulation, in which the stochastic Landau-Lifshitz-Gilbert equation is solved for each atom of a particle. The agreement was proven for simulations of a small GdFeCo ferrimagnetic particle with 70\,\% FeCo and 30\,\% Gd with a diameter of 5\,nm and a length of 10\,nm subject to various external applied fields and heat pulses.

In the last part of the work we investigated the difference between the ferriLLB equation and a more straightforward model of a ferrimagnet, in which the ferromagnetic sublattices are described with the ferroLLB and coupled with an intergrain exchange field. We derived this intergrain exchange field based on the Heisenberg Hamiltonian of the ferrimagnet under the assumption that the exchange is an interface exchange between the sublattices, with the interface extending across all atoms. The fact that both models produced identical results seemed surprising at first glance. But after comparing the individual field terms it turned out that almost every field term of the ferriLLB equation has a counterpart in the exchange coupled ferroLLB equations. For this reason, the good agreement can be well understood.

\section{Acknowledgements}
The authors would like to thank the Vienna Science and Technology Fund (WWTF) under grant No. MA14-044, the Advanced Storage Technology Consortium (ASTC), and the Austrian Science Fund (FWF) under grant No. I2214-N20 for financial support. The computational results presented have been achieved using the Vienna Scientific Cluster (VSC).\\

\appendix
\section{ferromagnetic LLB equation}
\label{sec:appendix}
The ferromagnetic LLB equation reads as follows~\cite{volger_llb}
\begin{eqnarray}
\label{eq:ferroLLB}
  \frac{d \boldsymbol{m}}{dt}= &-&\mu_0{\gamma'}\left( \boldsymbol{m}\times \boldsymbol{H}_{\mathrm{eff}}\right) \nonumber \\
  &-&\frac{\alpha_\perp\mu_0 {\gamma'}}{m^2} \left \{ \boldsymbol{m}\times \left [ \boldsymbol{m}\times \left (\boldsymbol{H}_{\mathrm{eff}}+\boldsymbol{\xi}_{\perp}  \right ) \right ] \right \}\nonumber \\
  &+&\frac{\alpha_\parallel  \mu_0{\gamma'}}{m^2}\boldsymbol{m}\left (\boldsymbol{m}\cdot\boldsymbol{H}_{\mathrm{eff}}  \right )+\boldsymbol{\xi}_{\parallel},
\end{eqnarray}
with $\gamma'=|\gamma_{\mathrm{e}}|/(1+\lambda^2)$. Longitudinal and perpendicular thermal field components consist of white noise random numbers with zero mean and variance
\begin{equation}
  \left \langle \xi_{\eta,i}(t,\boldsymbol{r})\xi_{\eta,j}({t}',\boldsymbol{r}') \right \rangle = 2D_\eta \delta_{ij}\delta(\boldsymbol{r}-\boldsymbol{r}')\delta(t-{t}').
\end{equation}
Diffusion constants $D_\eta$ can be computed per:
\begin{eqnarray}
  D_\perp&=&\frac{\left (\alpha_\perp-\alpha_\parallel  \right )k_{\mathrm{B}} T}{ \gamma' \mu^2_0 M_0 V \alpha^2_\perp}\nonumber\\
  D_\parallel&=&\frac{\alpha_\parallel \gamma' k_{\mathrm{B}} T}{M_0 V}.
\end{eqnarray}
The effective magnetic field consists of external field $\boldsymbol{H}_{\mathrm{ext}}$, anisotropy field
\begin{equation}
  \label{eq:HaniFerro}
   \mu_0\boldsymbol{H}_\mathrm{ani}=\frac{1}{\widetilde{\chi}_{\perp}}\left( m_x\boldsymbol{e}_{x}+m_y\boldsymbol{e}_{y}\right),
\end{equation}
and internal exchange field 
\begin{equation}
\label{eq:blochField}
 \mu_0\boldsymbol{H}_{\mathrm{J}}= \frac{1}{2\widetilde{\chi}_{\parallel}}\left( 1-\frac{m^2}{m^2_{\mathrm{e}}} \right)\boldsymbol{m} \quad\mathrm{for}\quad T\lesssim T_{\mathrm{C}}.
\end{equation}
Additionally, the intergrain exchange field of Eq.~\ref{eq:exchange_field}, as derived in Sec.~\ref{sec:Discussion} adds to the effective magnetic field.

%

\end{document}